\documentstyle[12pt]{article}
\begin{document}
\newcommand{\nc}{\newcommand}
\newcounter{romzahl}
\nc{\Rz}[1]{\setcounter{romzahl}{#1} \Roman{romzahl}}
\newtheorem{definition}{Definition}
\newtheorem{lemma}{Lemma}
\newtheorem{theorem}{Theorem}

\nc{\proof}{{\bf Proof: }}
\nc{\qed}{\mbox{$\;\clubsuit$}\par\hbox{}\par\noindent}

\nc{\para}{{\scriptscriptstyle\parallel}}
\nc{\senk}{{\scriptscriptstyle\perp}}

\nc{\be}{\begin{eqnarray}}
\nc{\ee}{\end{eqnarray}}
\nc{\nn}{\nonumber}
\nc{\ts}{\textstyle}

\nc{\mhspace}[1]{{\mbox{\hspace{#1}}}}
\nc{\mvspace}[1]{{\mbox{\vspace{#1}}}}
\nc{\mscript}[1]{_{\mbox{\scriptsize \it #1}}}

\nc{\ec}[3]{{C_{#1_#3}^{#2^#3_1#2^#3_2}}}
\nc{\er}[3]{R^{\, {#1_#3}}_{#2^#3_1#2^#3_2}}
\nc{\ecklamauf}[3]{C_{\,{#1_#3}}^{\{{#2^#3_1#2^#3_2}}}
\nc{\erklamauf}[3]{R^{#1_#3}_{\{{#2^#3_1#2^#3_2}}}
\nc{\ecklamzu}[3]{C_{#1_#3}^{#2^#3_1#2^#3_2\}_{as}}}
\nc{\erklamzu}[3]{R^{#1_#3}_{#2^#3_1#2^#3_2\}_{as}}}
\nc{\zc}[3]{\ec{#1}{#2}{1} \ldots \ec{#1}{#2}{#3}}
\nc{\zr}[3]{\er{#1}{#2}{1} \ldots \er{#1}{#2}{#3}}
\nc{\zcas}[3]{\ecklamauf{#1}{#2}{1} \ldots \ecklamzu{#1}{#2}{#3}}
\nc{\zras}[3]{\erklamauf{#1}{#2}{1} \ldots \erklamzu{#1}{#2}{#3}}
\nc{\gen}[1]{\mbox{\bf\sf #1}}
\nc{\funkop}[1]{{\cal #1}}
\nc{\frak}[1]{{\cal #1}}
\nc{\dalembert}{{\mbox{\large$\Box$}}}
\nc{\hatdalembert}{{\mbox{\large$\hat{\Box}$}}}
\nc{\christoffel}[2]{\left\{\!\!\begin{array}{c}{#1}\\
   {#2}\end{array}\!\!\right\}}

\nc{\ket}[1]{\left|\,{#1}\,\right>}
\nc{\bra}[1]{\left<\,{#1}\,\right|}
\nc{\braket}[2]{\left<\,{#1}\mid{#2}\,\right>}
\nc{\varket}[1]{\left|\,{#1}\,\right)}
\nc{\varbra}[1]{\left(\,{#1}\,\right|}
\nc{\varbraket}[2]{\left(\,{#1}\mid{#2}\,\right)}

\nc{\lek}{\left[}
\nc{\rek}{\right]}
\nc{\lrk}{\left(}
\nc{\rrk}{\right)}
\nc{\lgk}{\left\{}
\nc{\rgk}{\right\}}
\nc{\luk}{\left.}
\nc{\ruk}{\right.}
\nc{\klr}[1]{\left(\,#1\,\right)}
\nc{\klg}[1]{\left\{\,#1\,\right\}}
\nc{\kle}[1]{\left[\,#1\,\right]}
\nc{\dokle}[1]{\kle{{\mhspace{-6.5pt}}\kle{{#1}}{\mhspace{-6.5pt}}}}

%\nc{\ehoch}[1]{e^{^{\scriptstyle \; #1}}}
\nc{\ehoch}[1]{\exp \klg{#1}}
\nc{\dvier}{d^{^{\scriptstyle 4}} \!\!\! }
\nc{\ddrei}{d^{^{\scriptstyle \, 3}} \!\!\! }
\nc{\deltavier}{\delta^{^{\scriptstyle 4}} \!\! }

\nc{\SS}{{\cal S}}
\nc{\TT}{{\cal T}}
\nc{\HH}{{\cal H}}

\nc{\Sp}{S\!p\,}
\nc{\betr}[1]{\left\vert #1 \right\vert}
\nc{\norm}[1]{\left\Vert #1 \right\Vert}
\nc{\quer}[1]{\overline{#1}}
\nc{\matrixzz}[4]{\klr{\begin{array}{cc} #1 & #2 \\ #3 & #4
                  \end{array}}}
\nc{\ua}{\uparrow}
\nc{\da}{\downarrow}

\nc{\foda}[1]{\pi_{{\cal F}} \lrk \, #1 \, \rrk }
\nc{\dufoda}[1]{\pi_{{\cal F}^\ast } \lrk \, #1 \, \rrk }
\nc{\fovak}{\Omega _{{\cal F}}}
\nc{\dufovak}{\Omega _{{\cal F}^\ast }}
\nc{\fozu}[1]{\omega_{{\cal F}} \lrk \, #1 \, \rrk }
\nc{\dufozu}[1]{\omega_{{\cal F}^\ast } \lrk \, #1 \, \rrk }
\nc{\nfozu}{\omega_{{\cal F}}}
\nc{\ndufozu}{\omega_{{\cal F}^\ast}}
\nc{\omvw}[1]{\omega_ {v_{#1}w_{#1}}}

\nc{\hred}{H_{\mbox{\scriptsize red}}}
\nc{\heff}{H_{\mbox{\scriptsize eff}}}
\nc{\psibcs}{\psi_{_{BCS}}}

\nc{\ep}{\varepsilon}
\nc{\lam}{\lambda}
\nc{\sig}{\sigma}
\nc{\Lam}{\Lambda}
\nc{\om}{\omega}
\nc{\Om}{\Omega}
\nc{\al}{\alpha}
\nc{\ga}{\gamma}
\nc{\ka}{\kappa}
\nc{\vp}{\varphi}

\nc{\1}{{{\hbox{1}}\!{\hbox{l}}}}
\nc{\N}{{{\hbox{I}}\!{\hbox{N}}}}
\nc{\R}{{{\hbox{I}}\!{\hbox{R}}}}
\nc{\Z}{{{\hbox{Z}\!\!\hbox{Z}}}}
\nc{\C}{\hbox{\rlap{$\,\,$\hbox{\vrule height1.5ex width0.12ex
          depth0ex}}C}}
\nc{\Q}{{\hbox{\rlap{$\,\,$%
          \hbox{\vrule height6pt width1pt depth0.1pt}}Q}}}

\nc{\avonc}{{\cal A}\lrk \C \rrk }
\nc{\avoncn}{{\cal A}\lrk \C^n \rrk }
\nc{\aueinsvonc}{{{\cal A}_{U\lrk 1\rrk }} \lrk \C \rrk }
\nc{\avonczwei}{{\cal A}\lrk \C^2 \rrk }
\nc{\auzwei}{{{\cal A}_{U\lrk 2\rrk }} \lrk \C^2 \rrk }

\nc{\ct}{\tilde{c}}

\nc{\dm}{d\mu}
\nc{\dmf}{d\mu_{_{{\cal F}}}}
\nc{\dmfast}{d\mu_{_{{\cal F}^\ast}}}
\nc{\dxi}{d\xi}
\nc{\dxiast}{d\xi^{\ast}}
\nc{\dxixi}{d\xi^{\ast}d\xi}
\nc{\dxil}{d\xi^{}_l}
\nc{\dxilast}{d\xi_l^{\ast}}
\nc{\dxir}{d\xi^{}_r}
\nc{\dxirast}{d\xi_r^{\ast}}
\nc{\dxixil}{d\xi_l^{\ast}d\xi^{}_l}
\nc{\dxixir}{d\xi_r^{\ast}d\xi^{}_r}
\nc{\dxixilr}{d\xi_r^{\ast}d\xi^{}_rd\xi_l^{\ast}d\xi^{}_l}
\nc{\xiast}{\xi^{\ast}}
\nc{\xil}{\xi_l^{}}
\nc{\xilast}{\xi_l^{\ast}}
\nc{\xir}{\xi_r^{}}
\nc{\xirast}{\xi_r^{\ast}}
\nc{\klxi}{\klr{\xi,\,\xi^{\ast}}}
\nc{\klxilr}{\klr{\xil,\,\xilast,\,\xir,\,\xirast}}

\nc{\etal}{\eta_l^{}}
\nc{\etalast}{\eta_l^{\ast}}
\nc{\etar}{\eta_r^{}}
\nc{\etarast}{\eta_r^{\ast}}
\nc{\kletalr}{\klr{\etal,\,\etalast,\,\etar,\,\etarast}}

\nc{\cc}[3]{C^{{#1}{#2}}_{#3}}
\nc{\rr}[3]{R_{{#1}{#2}}^{\: #3}}
\nc{\vj}[2]{j_{I_1^{#1}}j_{I_2^{#1}}\ldots j_{I_1^{#2}}j_{I_2^{#2}}}
\nc{\abl}[1]{\frac{\delta }{\delta {#1}}}
\nc{\jabl}[1]{\frac{\delta }{\delta j_{I_{#1}}}}
\nc{\habl}[1]{\frac{\delta }{\delta h_{K_{#1}}}}
\nc{\zjabl}[1]{\frac{\delta }{\delta j_{I^{#1}_2}} \frac{\delta
}{\delta j_{I^{#1}_1}}}
\nc{\babl}[1]{\frac{\delta }{\delta b_{k_{#1}}}}
\nc{\bablk}[1]{\frac{\delta }{\delta b_{k{#1}}}}
\nc{\babll}[1]{\frac{\delta }{\delta b_{l{#1}}}}
\nc{\pabl}[1]{\frac{\partial }{\partial {#1}}}
\nc{\aket}[1]{\ket{{\cal A}\; [\:{#1}\:]}}
\nc{\bket}[1]{\ket{{\cal B}\; [\:{#1}\:]}}
\nc{\bbra}[1]{\bra{{\cal B}\; [\:{#1}\:]}}
\nc{\gbra}[1]{\bra{{\cal G}\; [\:{#1}\:]}}
\nc{\atildeket}[1]{\ket{\tilde{{\cal A}}\; [\:{#1}\:]}}
\nc{\tildefket}[1]{\ket{\tilde{{\cal F}}\; [\:{#1}\:]}}
\nc{\fket}[1]{\ket{{\cal F}\; [\:{#1}\:]}}
\nc{\tket}[1]{\ket{{\cal T}\; [\:{#1}\:]}}
\nc{\s}[2]{S^{#1_1\ldots #1_n}_{#2_1\ldots #2_n}}
\nc{\skl}{S^{k_1 \ldots k_n}_{l_1 \ldots \, l_n}}
\nc{\gammas}{\Gamma^{\, l \, l'}_{kk'} \, S^{m_1 \ldots m_n \: k
            \quad \: k'}_{m'_1 \ldots m'_n m'_{n+1} m'_{n+2}}}
\thispagestyle{empty}
\vspace*{2cm}
{\huge \bf \centerline{On quantum field theories with}
\centerline{finitely many degrees of freedom}}
\vspace*{2cm}
\centerline{R.~Kerschner}
\centerline{Institut f\"ur Theoretische Physik, University of T\"ubingen}
\centerline{Auf der Morgenstelle 14, 72076 T\"ubingen, Germany}
\vspace*{2cm}
{\Large \bf \centerline{Abstract}}
\begin{center}
\parbox{12cm}{The existence of inequivalent representations in
quantum field theory with
{\it finitely} many degrees of freedom is shown. Their properties
are exemplified and analysed for concrete and simple models. In
particular the relations to Bogoliubov--Valatin quasi-particles,
to thermo field dynamics, and to $q$--deformed quantum theories are
put foreward. The thermal properties of the non-trivial vacuum
are given and it is shown that the thermodynamic equilibrium
state is
uniquely obtained by an irreversible vacuum dynamics. Finally, the
theory is applied to a realistic model: the BCS--theory of
superconductivity. An exact solution in order $O(N^{-1})$ for
the full particle number conserving BCS--Hamiltonian with particle number
symmetric ground state is given.}
\end{center}

\newpage \setcounter{page}{1}
\tableofcontents

\section{Introduction}
The quantum theory of interacting many-particle systems is
governed by the use of field operators. In their Fock
representation these operators have the meaning of creating or
annihilating specific single-particle quantum states. For
quantum fields with
infinitely many degrees of freedom the technical
managing of the theory as well as its physical interpretation are
complicated by the appearance of infinitely many irreducible,
unitarily inequivalent representations. According to Haag's
theorem the Fock representation is not admissible for
interacting or self-interacting quantum field theories
\cite{haagtheorem}.
\nocite{haag55,barton,streaterwightman,friedrichs} As a
matter of principle, the use of the interaction picture and
perturbation theory are not allowed. In the
thermodynamic limit it is even impossible to obtain the thermal
properties of a free fermion or boson gas in the Fock
representation \cite{arakiww}. \nocite{arakiwoods63,arakiwyss64}
Thus, the need to go beyond the Fock
representation in many-particle theory is obvious.

The possibility of inequivalent representations must be
understood as the major difference between quantum mechanics and
quantum field theory. This structural enrichment of the theory
should be highly appreciated, since it provides the basis for
describing a large variety of correlations and collective
microscopic and macroscopic phenomena. Many rigorous approaches
have been put foreward in axiomatic and algebraic quantum
theory, showing the relevance of inequivalent representations to
concrete physical problems \cite{brattelirobinson,idaanw}.
\nocite{emch,gardingwightman54a,rieckersullrich85b,rieckers90}
\nocite{gueninmisra63,dellantoniodoplicher67,kmlm72,morchiostrocchi80}
\nocite{sewell73,bona88,fkv91,fnw89,gvv90}
However, the mathematical
effort to achieve the results is rather high.

Due to von Neumann there are --- unfortunately --- no unitarily
inequivalent,
irreducible representations of the canonical commutation or
anticommutation relations for quantum field operators with
finitely many degrees of freedom. From the physical point of
view this seems to be
unsatisfactory because the physics of an arbitrarily large but
finite system differs remarkably from its infinite
limit. In addition, explicit calculations could be simplified,
if the appearance of inequivalent representations were already given
in quantum theories with finitely many degrees of freedom.

Since it is impossible to circumvent von Neumann's theorem let
us ask: `How do inequivalent representations come about in
taking the infinite volume limit?'. Consider field operators
$c_k$ and $c^\ast_k$ and an extensive observable such as the
particle number operator $N$. For {\it finitely\/} many degrees of freedom
the particle number operator is given by
\be
N = \sum_k c^\ast_k c_k
\label{eq1}
\ee
in its Fock representation.
Going over to the infinite system expression (\ref{eq1}) becomes
undefined, i.e. the particle number operator does not belong to
the algebra generated by the field operators any longer. Rather, one has to
pass to the von Neumann algebra which is the weak closure of the
field operator algebra in order to get the particle number
operator as a well defined object \cite{brattelirobinson,idaanw}.
On the other hand the commutator
\be
\kle{N,\,c_k} = - c_k
\label{eq2}
\ee
is not affected by taking the infinite volume limit.

Therefore a
unified quantum field theory of finite and infinite systems
should be constructed from steady expressions such as
(\ref{eq2}) alone. This enforces the use of an enlarged algebra
of field operators and observables, which --- as will be shown
--- possesses inequivalent
representations for finitely many degrees of freedom.

\section{Quantum field theory with finitely many degrees of freedom}

\subsection{Fundamentals}
Let us start with the well known CAR--algebra ${\cal A}(\HH)$
defined by
\be
\klg{c\klr{f},\,c\klr{h}} = 0\, , \quad
\klg{c^\ast\klr{f},\,c\klr{h}} =  \braket{f}{h} \1
\label{eq3}
\ee
(and $f\rightarrow c(f)$ is anti-linear),
where $f,h$ are elements of a finite dimensional Hilbert space
$\HH=\C^n$.\footnote{The restriction to the CAR--algebra is not
essential, but was chosen here because it allows explicite matrix
representations for the examples discussed below.} In addition
to the algebra of field operators we have a symmetry group with
the generators describing the observables of the theory.
These observables and their relations to the
CAR--algebra have to be fixed, too. For a symmetry group
$G\subseteq U(n)$ there is a unitary representation $u_g$ for
$g\in G$ on $\HH=\C^n$ defining a $\ast$--automorphism $\al_g$ on the
CAR--algebra in terms of
\be
\al_g \klr{c\klr{f}} = c\klr{u_g f} \;.
\label{eq4}
\ee
We require this $\ast$--automorphism to be unitarily implemented
on the CAR--algebra.
\begin{definition}
The group $G$ and the $\ast$--automorphism
(\ref{eq4}) are said to be unitarily implemented
on the CAR--algebra, if to any $g\in G$ there corresponds an
operator $U_g$ with {\rm
\be
U_e =\1\,,\quad U_{g_1g_2} = U_{g_1}U_{g_2}\,,\quad U_{g^{-1}} =
U^\ast_g
\label{eq5}
\ee}
and
\be
\al_g\klr{A} = U_g A U^\ast_g
\label{eq6}
\ee
for all $A\in {\cal A} (\HH)$.
\end{definition}
It follows from (\ref{eq5}) that $U_g U^\ast_g = U^\ast_g U_g
=\1$. Furthermore,
$U_gAU^\ast_g$ is always an element of the CAR--algebra, but
this need not be true for $U_g$. The algebra generated by the
field operators and the independent operators $U_g$ is in fact
an (infinite dimensional) extension of the
CAR--algebra and will be denoted by ${\cal A}_G(\HH)$.

The calculation of expectation values, transition amplitudes,
etc. requires the concept of states. In algebraic
quantum theory these are introduced as linear, normalized,
positive \C--valued functionals on the operator algebra under
consideration. We shall concentrate on vacuum states here.
\begin{definition}
A vacuum state on the algebra ${\cal A}_G(\HH)$ is a linear,
normalized, positive functional $\om$ with the additional property
\be
\om\klr{A} = \om \klr{AU_g}
\label{eq7}
\ee
for all $A\in {\cal A}_G(\HH)$ and all $g\in G$.
\end{definition}
It follows from (\ref{eq7}) that $\om\klr{A} = \om \klr{U_gA}$.
The set of vacuum states is convex, i.e. any convex combination
of two vacuum states is again a vacuum state.
Sometimes a vacuum state is called $G$--invariant, too.

If compared with the Wightman axioms \cite{wightman56} of
quantum field theory there
show up to be strong interrelations if the states are visualized
by their GNS--representation
$\om(A)=\braket{\Om_\om}{\pi_\om(A)\,\Om_\om}$.
The main difference is that
the symmetry group $G\subseteq U(n)$ is much simpler than the Poincar\'e
group. Therefore we can drop the restrictions on the domains of
definition here. In addition we take a different point of view
concerning the uniqueness of the vacuum. The vacuum is not
determined uniquely by the symmetry properties (\ref{eq7})
alone, but requires the explicite dynamics of a concrete model.
However, besides of these points the following examples can be viewed as
realisations of the Wightman axioms for simple symmetry groups.

Since we will evaluate the theory only for the Lie groups
$G=U(n)$ we can put down
the fundamental equations in terms of the generators as well.
Denoting $c_\al=c\klr{f_\al}$ for a fixed orthonormal basis
$f_\al$ of $\HH$ and the generators of $u_g$ and $U_g$ by $q^k$
and $Q^k= (Q^k)^\ast$, respectively, we get the algebraic
relations
\be
\klg{c_\al,\,c_\beta}=0\,,\quad \klg{c^\ast_\al,\,c_\beta}=
\delta_{\al\beta} \1 \nn \\[1ex]
\kle{Q^k,\,c_\al} = \sum_\beta q^k_{\al\beta} \, c_\beta
\,,\quad  \kle{Q^k,\,c^\ast_\al} = \sum_\beta
\hat{q}^k_{\al\beta} \, c^\ast_\beta
\label{eq8}
\ee
with $\hat{q}^k_{\al\beta}= -\bar{q}^k_{\al\beta}$, (the bar
denotes complex conjugation). Furthermore, the generators satisfy
the Lie relations
\be
\kle{Q^k,\, Q^l} = i\,{f^{kl}}_m\, Q^m \,,\quad
\kle{q^k,\, q^l} = i\,{f^{kl}}_m\, q^m
\label{eq9}
\ee
with the structure constants ${f^{kl}}_m$ of the corresponding
Lie algebra. Vacuum states are characterized by the equations
\be
\om \klr{A\,Q^k} =0
\label{eq10}
\ee
for any $A\in {\cal A}_G(\HH)$ and all observables $Q^k$. Again
it follows that $\om \klr{Q^kA} =0$, too.

We can also introduce the notion of eigenvalue equations in this
formulation, which will prove to be useful for concrete calculations.
\begin{definition}
An operator $B_q$ is called (right) eigenvector of $Q$ with
(right) eigenvalue $q$, if $\om\klr{B^\ast_qB_q}\not=0$ and
\be
\om\klr{A\,Q\,B_q} = q\,\om\klr{A\,B_q}
\label{eq11}
\ee
for all $A\in {\cal A}_G(\HH)$.
\end{definition}
The definition of left eigenvectors and left eigenvalues is
obvious. One can show that left and right eigenvalues coincide
and are real for Hermitean observables $Q=Q^\ast$. Further, if
$B_q$ is a right eigenvector then $B^\ast_q$ is a left eigenvector
to the same eigenvalue.

\subsection{Examples}
We will now evaluate the previous setup for the
cases $\HH=\C$ and $\HH=\C^2$ and thereby give explicite
examples for inequivalent vacuum representations in quantum
field theory with finitely many degrees of freedom.

For $\HH =\C$ and $G=U(1)$ the field operator and observable
algebra is characterized by the equations
\be
\klg{c,\,c}=0\,,\quad \klg{c^\ast,\,c}=\1\,, \quad \kle{N,\,c} =
-c\,,
\label{eq12}
\ee
with $N$ being the Hermitean generator of the unitarily
implemented $U(1)$ symmetry.

A state on this algebra is
completely fixed, if its values on all algebraically independent
elements are known. By making use of the commutation relations
(\ref{eq12}) any algebra element can be `normal-ordered' such
that all operators $N$ are to the right of all field operators.
Taking these normal-ordered operators as algebraic basis
requires any vacuum state to be zero on all elements except the
CAR--algebra of field operators. Hence, any vacuum state is
completely determined by its values on the CAR--algebra.

\begin{lemma}
The set of all vacuum states on the algebra defined in (\ref{eq12})
is parametrized by the real parameter $v=\om\klr{cc^\ast}$,
with $v\in\kle{0,1}$.
\end{lemma}
\proof
An algebraic basis of the CAR--algebra is given by  ${\cal
B}=\klg{\1, \,c, \,c^\ast, \,cc^\ast}$, whose elements satisfy
the eigenvalue equations
\be
\om\klr{AN\1}= 0\,,\quad \om\klr{ANcc^\ast}=0 \nn\\[1ex]
\om\klr{ANc}=-\om\klr{Ac}\,,\quad \om\klr{ANc^\ast}=\om\klr{Ac^\ast}\;.
\label{eq13}
\ee
Setting $A=\1$ yields $\om(c)=\om(c^\ast)=0$, and together with
the normalization
$\om\klr{\1}=1$ the only undetermined value is $\om(cc^\ast)=v$.
{}From the positivity of $\om$ it follows that $v\geq 0$. With the aid
of (\ref{eq12}) and the linearity of $\om$ we get
$\om\klr{c^\ast c}=1-v$ and positivity shows that $1-v\geq 0$.
\qed
Labeling the states in terms of their values on $cc^\ast$ we can
give their extremal decomposition in the following way:
\begin{theorem}
The convex set of vacuum states $\klg{\om_v;v\in\kle{0,1}}$ possesses
two extremal, pure states $\om_1$ and $\om_0$. The extremal
decomposition of $\om_v$ is given by
\be
\om_v =v\,\om_1 + \klr{1-v}\om_0 \;.
\label{eq14}
\ee
\end{theorem}
\proof
Assume that $\om_1 $ had some non-trivial decomposition $\om_1 =
\lam\, \om_{v_1} + \klr{1-\lam} \om_{v_2}$ with $0<\lam<1$.
Applied to $c^\ast c$ we get
$\lam\klr{1-v_1}+\klr{1-\lam}\klr{1-v_2}=0$. Since $\lam$ and
$1-\lam$ are positive it follows that $v_1=v_2=1$. Therefore
$\om_1$ can only be decomposed trivially. In the same way one
shows that $\om_0$ is extremal. It remains to show that the
extremal decomposition (\ref{eq14}) is complete. This follows,
if (\ref{eq14}) is evaluated on an arbitrary algebra element. It
suffices to consider normal-ordered algebra elements upon which
$\om_v $ does not vanish identically, i.e.
$A=\al_1\1+\al_2cc^\ast$. We get
$v\om_1(A)+(1-v)\om_0(A)=v(\al_1+\al_2)+(1-v)\al_1 = \al_1 +
v\al_2 =\om_v(A)$.
\qed
The convex set of vacua consists of the line segment between the
extremal vacua $\om_1$ and $\om_0$. Hence, it has a simplex structure!
This is a rather unusual feature in ordinary quantum theory, but
it is the characteristic state space structure whenever
inequivalent representations occur \cite{brattelirobinson}.

Explicite matrix representations have to
be reconstructed from the vacuum states. Due to the
extremal decomposition (\ref{eq14}) the general representation
is reducible and the irreducible components are given by the
representations belonging to the extremal states. For the
formal GNS--construction
\be
\om_v\klr{A} = \bra{\Om_v} \pi_v(A) \ket{\Om_v}
\label{eq15}
\ee
it must be
\be
\pi_v(A) = \pi_1(A)\oplus\pi_0(A) \,,\quad \ket{\Om_v} =
e^{i\al}\sqrt{v} \ket{\Om_1} \oplus \sqrt{1-v} \ket{\Om_0}
\label{eq16}
\ee
with an undetermined relative phase factor $e^{i\al}$.

The extremal representations $\pi_1$ and $\pi_0$ along with their
extremal vacuum vectors $\ket{\Om_1}$ and
$\ket{\Om_0}$ can be reconstructed from (\ref{eq15}) for $v=1$ and
$v=0$, respectively, by taking into consideration that
$\klg{\ket{\Om_1}, \pi_1(c^\ast)\ket{\Om_1}}$ and
$\klg{\ket{\Om_0}, \pi_0(c)\ket{\Om_0}}$ are orthonormal basis
systems for $v=1$ and $v=0$, respectively. One gets the matrix
representations
\be
\pi_1(c) = \pi_0(c) = \klr{\begin{array}{cc} 0&0\\1&0\end{array}}\;, \nn\\[1ex]
\pi_1(N) = \klr{\begin{array}{cc} 1&0\\0&0\end{array}}\,,\quad
\pi_0(N) = \klr{\begin{array}{cc} 0&0\\0&-1\end{array}} \;,
\label{eq17}
\ee
and
\be
\ket{\Om_1} = \klr{\begin{array}{c} 0\\1\end{array}}\,,\quad
\ket{\Om_0} = \klr{\begin{array}{c} 1\\0\end{array}} \;.
\label{eq18}
\ee
Next to the Fock representation $\pi_1$ with Fock vacuum
$\ket{\Om_1}$ we have the extremal representation $\pi_0$ which we
shall call the dual Fock representation, because it is
characterized by $\pi_0(c^\ast)\ket{\Om_0} =0$.
Obviously still von Neumann's theorem is not falsified here,
since the irreducible representations of the CAR--algebra
alone are equivalent. However the larger field operator and
observable algebra ${\cal A}_G(\HH)$ does indeed have
irreducible, inequivalent, and finite dimensional
representations $\pi_1$ and $\pi_0$,
because of the inequivalent spectrum of the particle number
observable $N$ in both representations.

For the extremal vacua
the particle number operator can be expressed as
$\pi_1(N)=\pi_1(c^\ast c)$ or $\pi_0(N) = \pi_0 (c^\ast c -\1)$,
respectively,
but aside these extremal cases the particle number operator does
not belong to the CAR--algebra. Like in the case of infinitely
many degrees of freedom the observables do not belong to the
CAR--algebra in general vacuum representations.

If only vacuum representations
are taken into consideration an arbitrary element of ${\cal
A}_G(\HH)$ can be written in the form $A=A_1+NA_2$ with $A_1,
A_2$ being elements of the CAR--algebra.

Our results might appear to be a little hair-splitting, because the
Fock and dual Fock representations could be converted into one
another by reinterpreting particles as holes and vice versa and
renormalizing the particle number operator. In passing over to
the more complicated case $\HH=\C^2$ and $G=U(2)$ however, we
will find a third extremal representation which is completely
different from the Fock and dual Fock representations.

We seek to find all irreducible vacuum representations of the
(anti)commutation relations
\be
\klg{c_\al,\,c_\beta}=0\,,\quad \klg{c^\ast_\al,\,c_\beta}=
\delta_{\al\beta} \1 \,,\quad \kle{N,\,c_\al} = - c_\al \nn \\[1ex]
\kle{S^k,\,c_\al} = \sum_\beta
\sig^k_{\al\beta} \, c_\beta  \,,\quad
\kle{S^k,\, S^l} = i\,\ep_{klm}\, S^m \,, \quad \kle{N,\,S^k}=0\;,
\label{eq19}
\ee
where $\al,\beta \in\klg{1,2}$, $S^k$ are the implemented spin
operators and $\sig^k$ the Pauli spin matrices.

Calculating the
eigenvalues of $N,S^3$ and $\vec{S}^{\,2}$ we get the following
tabular of simultaneous eigenvectors and their
eigenvalues\footnote{It is
possible that some of the formal eigenvectors are zero vectors.
This will depend on their representation.}
\begin{center}
\begin{tabular}{|c|c|c|c|}
\hline
eigenvector &
   \rule{12pt}{0pt} $S_3$ \rule{12pt}{0pt} &
   \rule{14pt}{0pt} $\vec{S}^2$ \rule{12pt}{0pt} &
   \rule{12pt}{0pt} $N$ \rule{12pt}{0pt} \\[0.2ex]
\hline
%%% & & & \\
$ \1    $ & $   0   $ & $   0   $ & $   0  $\\[0.2ex]
$\frac{1}{2}\klr{ c^{}_1 c_1^\ast + c^{}_2 c_2^\ast}$&$0$&$0$&$0$\\[0.2ex]
$c^{}_1 c^{}_2 c_2^\ast c_1^\ast         $ & $ 0 $ & $ 0 $ & $ 0$\\[0.2ex]
$c^{}_1 c_2^\ast                      $ & $ 1 $ & $ 2 $ & $ 0$\\[0.2ex]
$\frac{1}{2}\klr{ c^{}_1 c_1^\ast - c^{}_2 c_2^\ast}$&$0$&$2$&$0$\\[0.2ex]
$c^{}_2 c_1^\ast  $ & $-1 $ & $ 2 $ & $ 0$\\[0.2ex]
$c^{}_1           $ & $ \frac{1}{2} $&$\frac{3}{4}$&$-1$\\[0.2ex]
$c^{}_1 c^{}_2 c_2^\ast $ & $ \frac{1}{2} $&$\frac{3}{4}$&$-1$\\[0.2ex]
$c^{}_2              $ & $ -\frac{1}{2} $&$\frac{3}{4}$&$-1$\\[0.2ex]
$c^{}_2 c^{}_1 c_1^\ast $ & $ -\frac{1}{2} $&$\frac{3}{4}$&$-1$\\[0.2ex]
$c_2^\ast         $ & $ \frac{1}{2} $&$\frac{3}{4}$&$1$\\[0.2ex]
$c^{}_1 c_1^\ast c_2^\ast$&$\frac{1}{2}$&$\frac{3}{4}$&$1$\\[0.2ex]
$c_1^\ast         $&$-\frac{1}{2}$&$\frac{3}{4}$&$1$\\[0.2ex]
$c^{}_2 c_2^\ast c_1^\ast$&$-\frac{1}{2}$&$\frac{3}{4}$&$1$\\[0.2ex]
$c^{}_1 c^{}_2             $ & $  0  $ & $  0  $ & $ - 2 $\\[0.2ex]
$c_1^\ast c_2^\ast   $ & $  0  $ & $  0  $ & $ 2 $\\[0.2ex]
\hline
\end{tabular} .
\end{center}
Although the Pauli principle is fully valid, there appears to be
a spin--1 triplet of eigenvectors. However, these are zero
vectors in the Fock
and dual Fock representations, inspiring one to search
for representations where they are non-zero.

We shall systematically evaluate all vacuum representations of
the relations (\ref{eq19}). By the same arguments as before any
vacuum state is completely determined by its values on an
algebraic basis of the
CAR--algebra, which is chosen to be the set of eigenvectors in
the tabular above. Like in (\ref{eq13}) we conclude that a
vacuum state is zero on all eigenvectors unless all its
eigenvalues are zero. Therefore we have:
\begin{lemma}
A vacuum state on the algebra given by (\ref{eq19}) is
completely characterized in terms of two real parameters
$v=\om_{vw}(c_1c_1^\ast) = \om_{vw}(c_2c_2^\ast)$ and
$w=\om_{vw}(c_1c_2c_2^\ast c_1^\ast)$ restricted by the
inequalities $1\geq v\geq w\geq 0$ and $w\geq 2v-1$.
\end{lemma}
\proof
Enumerating the eigenvectors by $e_1=\1, \, e_2={\ts
\frac{1}{2}} \klr{c_1c_1^\ast+c_2c_2^\ast},\, \ldots,\, e_{16} =
c_1^\ast c_2^\ast$ the restrictions on the parameters $v$ and $w$
follow from the positivity of $\om(e_i^\ast e_i)\geq 0$ for
$i\in\klg{1,2,\ldots,16}$.
\qed
\begin{theorem}
There are three extremal vacuum states $\om_{11}$, $\om_{00}$
and $\om_{\frac{1}{2}0}$. The extremal decomposition of a
general vacuum state is given by
\be
\om_{vw}= w \, \om_{11} + \klr{1-2v+w} \om_{00} +2\klr{v-w}
\om_{\frac{1}{2}0}
\label{eq20}
\ee
\end{theorem}
\proof
Here $\om_{11}$ and $\om_{00}$ are the Fock and dual Fock vacua,
respectively. We shall only prove explicitely that
$\om_{\frac{1}{2}0} $ is extremal. Suppose there were a
non-trivial decomposition $\om_{\frac{1}{2}0} = \lam\om_{v_1w_1}
+ (1-\lam)\om_{v_2w_2}$ with $\lam \in ]0,1[$. Evaluated on
$c_1c_2c^\ast_2c^\ast_1$ gives $0=\lam w_1+(1-\lam)w_2$, implying
$w_1=w_2=0$. Applying the decomposition to
$c^\ast_1c^\ast_2c_2c_1$ and using $w_1=w_2=0$ yields
$0=\lam(1-2v_1)+ (1-\lam)(1-2v_2)$. Therefore
$v_1=v_2=\frac{1}{2}$ and $\om_{\frac{1}{2}0} $ is extremal. In
order to determine the decomposition (\ref{eq20}) we evaluate
the ansatz
$\om_{vw}=\lam_1\om_{11}+\lam_2\om_{00}+\lam_3\om_{\frac{1}{2}0}$
for the algebra elements $\1,\, c_1c^\ast_1$ and
$c_1c_2c^\ast_2c^\ast_1$. The resulting equations
$1=\lam_1+\lam_2+\lam_3,\, v=\lam_1+\frac{1}{2}\lam_3$ and
$w=\lam_1$ can be resolved to give (\ref{eq20}).
\qed
Again the corresponding vacuum representation decomposes into a
direct sum of the extremal representations
\be
\pi_{vw}=\pi_{11}\oplus \pi_{00} \oplus \pi_{\frac{1}{2}0} \;,
\label{eq21}
\ee
where the vacuum vector
\be
\ket{\Om_{vw}} = e^{i\al_1}\sqrt{w}\ket{\Om_{11}} \oplus
e^{i\al_2} \sqrt{1-2v+w} \ket{\Om_{00}} \oplus \sqrt{2(v-w)}
\ket{\Om_{\frac{1}{2}0}}
\label{eq22}
\ee
(with two arbitrary relative phases) describes the mixing of the
extremal representations. One finds that $\pi_{11}$ and
$\ket{\Om_{11}}$ are the Fock representation and Fock vacuum.
Similar $\pi_{00}$ and $\ket{\Om_{00}}$ are the dual Fock
representation and dual Fock vacuum, completely characterized by
the equations $\pi_{00}(c^\ast_1) \ket{\Om_{00}} =
\pi_{00}(c^\ast_2) \ket{\Om_{00}} =  0$. Since these
representations are well known, we shall concentrate on the
extremal representation $\pi_{\frac{1}{2}0}$ with vacuum
$\ket{\Om_{\frac{1}{2}0}}$.
\begin{lemma}
The Fock and dual Fock representations are both irreducible
$4\times4$ matrix representations, containing spin--0 and
spin--1/2 configurations with particle numbers 0, 1, or
2 for the Fock and 0, -1, or -2 for the dual Fock representation,
respectively.
\end{lemma}
\proof The proof is trivial and omitted here.
\qed
Lemma \thelemma\ must be contrasted with the following result:
\begin{lemma}
The extremal representation $\pi_{\frac{1}{2}0}$ is an
irreducible, full $8\times8$ matrix representation. There are
spin--0, spin--1/2 and spin--1 configurations. The
particle number can take the values 0, 1 and -1.
\end{lemma}
\proof
In analysing the scalar products $\om_{vw}(e^\ast_ie^{}_j)$ of
the algebraic basis of eigenvectors one finds that for
$(v,w)=(\frac{1}{2},0)$ a complete orthonormal basis is given by
the set $\{\beta_1=\1 ,\,\beta_2=\sqrt{2} \,c^{}_1 c_2^\ast,\,
\beta_3=c^{}_1 c_1^\ast - c^{}_2 c_2^\ast,\,
\beta_4=\sqrt{2} \; c^{}_2 c_1^\ast,\,
\beta _5=\sqrt{2} \, c^{}_1,\, \beta_6=\sqrt{2} \, c^{}_2,\,
\beta_7=\sqrt{2} \, c_1^\ast,\, \beta_8=\sqrt{2} \, c_2^\ast
\}$, i.e. $\om_{\frac{1}{2}0}(\beta_i^\ast
\beta_j)=\delta_{ij}$. Making use of this basis we get explicite
matrix representations of the field operators and observables by
evaluating
\be
\pi^{ij}_{\frac{1}{2}0}(A) = \om_{\frac{1}{2}0} (\beta^\ast_i A
\beta_j)\;.
\label{eq23}
\ee
It follows that
\be
\pi_{\frac{1}{2}0} \klr{c^{}_1} & = & \klr{\begin{array}{cccccccc}
                             0 & 0 & 0 & 0 & 0 & 0 &
                             \frac{1}{\sqrt{2}} & 0 \\
                             0 & 0 & 0 & 0 & 0 & 0 & 0 & 1 \\
                             0 & 0 & 0 & 0 & 0 & 0 &
                             \frac{1}{\sqrt{2}} & 0 \\
                             0 & 0 & 0 & 0 & 0 & 0 & 0 & 0 \\
                             \frac{1}{\sqrt{2}} & 0 &
                             \frac{-1}{\sqrt{2}} & 0 & 0 & 0 & 0
                             & 0 \\
                             0 & 0 & 0 & -1 & 0 & 0 & 0 & 0 \\
                             0 & 0 & 0 & 0 & 0 & 0 & 0 & 0 \\
                             0 & 0 & 0 & 0 & 0 & 0 & 0 & 0
                             \end{array}}
\nonumber \\[2ex]
\pi_{\frac{1}{2}0} \klr{c^{}_2} & = & \klr{\begin{array}{cccccccc}
                             0 & 0 & 0 & 0 & 0 & 0 & 0 &
                             \frac{1}{\sqrt{2}} \\
                             0 & 0 & 0 & 0 & 0 & 0 & 0 & 0 \\
                             0 & 0 & 0 & 0 & 0 & 0 & 0 &
                             \frac{-1}{\sqrt{2}} \\
                             0 & 0 & 0 & 0 & 0 & 0 & 1 & 0 \\
                             0 & -1 & 0 & 0 & 0 & 0 & 0 & 0 \\
                             \frac{1}{\sqrt{2}} & 0 &
                             \frac{1}{\sqrt{2}} & 0 & 0 & 0 & 0
                             & 0 \\
                             0 & 0 & 0 & 0 & 0 & 0 & 0 & 0 \\
                             0 & 0 & 0 & 0 & 0 & 0 & 0 & 0
                             \end{array}}
\label{eq24}
\ee
for the field operators and
\be
\pi_{\frac{1}{2}0} (S_k) = 0  \oplus  \Sigma_k  \oplus
\sigma_k  \oplus  \hat{\sigma}_k
\label{eq25}
\ee
with the one dimensional zero matrix $0$ and
\be
\Sigma_1 = \klr{\begin{array}{ccc}
                0 & \frac{-1}{\sqrt{2}} & 0 \\
                \frac{-1}{\sqrt{2}} & 0 & \frac{1}{\sqrt{2}} \\
                0 & \frac{1}{\sqrt{2}} & 0
                \end{array}}, \;
\Sigma_2 = \klr{\begin{array}{ccc}
                0 & \frac{i}{\sqrt{2}} & 0 \\
                \frac{-i}{\sqrt{2}} & 0 & \frac{-i}{\sqrt{2}} \\
                0 & \frac{i}{\sqrt{2}} & 0
                \end{array}}, \;
\Sigma_3 = \klr{\begin{array}{ccc}
                1 & 0 & 0 \\
                0 & 0 & 0 \\
                0 & 0 & -1
                \end{array}}\;.
\label{eq26}
\ee
Further the observables $N$ and $\vec{S}^{\,2}$ are diagonal
with $\pi_{\frac{1}{2}0}(N) = \mbox{diag}(0,0,0,0,-1,-1,1,1)$
and $\pi_{\frac{1}{2}0} (\vec{S}^{\,2} ) =
\mbox{diag}(0,2,2,2,3/4,3/4,3/4,3/4)$. Finally the vacuum vector
is represented by $\Om_{\frac{1}{2}0} = (1,0,0,0,0,0,0,0)$.
\qed
Notice that the representation (\ref{eq24}) of the CAR--algebra
is reducible. Analysing the characters shows it to be unitarily
equivalent to a direct sum of two Fock representations.
Nevertheless the representation $\pi_{\frac{1}{2}0}$ of the
complete algebra ${\cal A}_G(\HH)$ is the full $8\times8$ matrix
algebra and completely different in its structure from the Fock
or dual Fock representations. It can be shown that any algebra
element in the representation $\pi_{\frac{1}{2}0}$ is of the
form $A=A_0+\sum_{k=1}^3 S_k A_k$ with $A_0,\ldots,A_3$ being
elements of the CAR--algebra.
\begin{lemma}
The representation $\pi_{\frac{1}{2}0}$ is
characterized by the equations
\be
\pi_{vw} (c_1c_2) \ket{\Om_{vw}} =
\pi_{vw} (c^\ast_1c^\ast_2) \ket{\Om_{vw}}
=0 \;.
\label{eq27}
\ee
\end{lemma}
\proof
If $\pi_{vw} (c^\ast_1c^\ast_2)
\ket{\Om_{vw}} =0$ it follows that $\om_{vw}
(c_2c_1c^\ast_1c^\ast_2) = w = 0$. Similar we have
$\om_{vw}(c^\ast_1c^\ast_2c_2c_1)= 1-2v+w =0$. Hence $\om_{vw}=
\om_{\frac{1}{2}0}$.
\qed
\section{Quasi-particles and thermodynamics in Fock space}

In the previous section it was shown that inequivalent
representations might well exist in quantum theory with finitely
many degrees of freedom. In order to get some more
insight into their physical meaning we will compare these
results to different techniques in many-particle theory.

\subsection{The Bogoliubov--Valatin transformation}

The quasi-particle transformation introduced by Bogoliubov
\cite{bogoliubov58} and Valatin \cite{valatin58} is used to
describe BCS--like correlations in the theory of
superconductivity, nuclear theory, etc. in Fock space. It is
given in terms of the quasi-particle operators
\be
\al_1 = \sqrt{1-h}\,c_1-\sqrt{h}\,c^\ast_2\,,\quad \al_2 =
\sqrt{h}\,c^\ast_1 + \sqrt{1-h}\, c_2
\label{eq28}
\ee
in Fock representation. The `vacuum' of the quasi-particles
defined by $\al_\sig\ket{BV} =0$ follows as
\be
\ket{BV} = \klr{\sqrt{1-h} + \sqrt{h}\, c^\ast_1c^\ast_2} \ket{0} \;,
\label{eq29}
\ee
with $\ket{0}$ being the Fock vacuum ($c_\sig\ket{0}=0$).

If the Bogoliubov--Valatin state is defined by
\be
\om^{BV}_h(A) = \bra{BV(h)}A\ket{BV(h)}
\label{eq30}
\ee
one calculates that it takes exactly the same values on the
CAR--algebra
as the vacuum state $\om_{1-h,1-h}= (1-h)\om_{11} + h\,\om_{00}$
with the exception of the basis elements $c_1c_2$ and
$c^\ast_1c^\ast_2$.

This was to be expected, because the
Bogoliubov--Valatin transformation does not conserve the
particle number symmetry and we required our vacuum states to be
exact vacua with respect to the quantum numbers of observables.
In addition the extremal vacuum vectors $\ket{\Om_{11}}$ and
$\ket{\Om_{00}}$ have exactly the same properties as $\ket{0}$
and $c^\ast_1c^\ast_2\ket{0}$, respectively, except the fact
that $c^\ast_1c^\ast_2\ket{0}$ has particle number 2 instead of
0. Contrary to the mixing of creation and annihilation operators
(\ref{eq28}) in their Fock representation one can mix the Fock
and dual Fock representations of the original field operators
and simultaneously have an exact particle number symmetric
vacuum here.

Still, not every vacuum representation may be obtained by a
Bogoliubov--Valatin transformation. No contributions from the
extremal representation
$\pi_{\frac{1}{2}0}$ can be simulated by Bogoliubov--Valatin
quasi-particles. However, we will show below that this is possible by
introducing more general quasi-particles such as $q$--deformed
quantum fields in their Fock representation.

\subsection{$q$--deformed CAR--algebras}

In modern models of condensed matter and nuclear physics many
attempts have been made to describe correlated many-particle
quantum systems in terms of deformed quantum fields and deformed
symmetries \cite{qgrappl}.
\nocite{biedenharntarlini90,bonatsos92,bdrrs91}
\nocite{daskaloyannis92,floratos90,fradkin,macfarlane89,rrs90}
Here we will simulate the effects of inequivalent
representations of the algebras ${\cal A}_G(\HH)$ in the Fock
representation of the corresponding $q$--deformed CAR--algebras. We
want to concentrate on the case $\HH=\C^2$
showing how $q$--deformed CAR--algebras go beyond
Bogoliubov--Valatin quasi-particles and at the same time belong
to the general setup of vacuum representations on ${\cal
A}_G(\HH)$.

The $q$--deformed CAR--algebra will be characterized by the
relations
\be
a^{\dag}_{\al} a^{}_\al = \kle{N^{}_\al}\,,\quad
a^{}_\al a_{\al}^{\dag} = \kle{N^{}_\al + 1}
\nonumber \\
\klg{a^{\mbox{}}_1, a^{}_2} = \klg{a^{}_1, a_{2}^{\dag}} =
\klg{a_{1}^{\dag}, a^{}_2} = \klg{a_{1}^{\dag},
a_{2}^{\dag}} = 0
\nn\\
\kle{N^{}_\al, a^{}_\beta} = -\delta^{}_{\al \beta} \;
a^{}_\beta\,,\quad
\kle{N^{}_\al, a_{\beta}^{\dag}} = \delta^{}_{\al \beta} \;
a_{\beta}^{\dag}
\label{eq37}
\ee
with the deformation function
\be
\kle{N}=\frac{q^N-q^{-N}}{q-q^{-1}}
\label{eq38}
\ee
and the $SU_q(2)$ generators are
\be
J_+=a^{\dag}_1a_2\,,\quad J_-=a^{\dag}_2a_1\,,\quad J_3=
\frac{N_1-N_2}{2} \;.
\label{eq39}
\ee
Putting down the Fock representation by requiring
\be
a_1 \ket{0,0} = a_2 \ket{0,0} = 0
\label{eq40}
\ee
we can use the normalized set of state vectors
\be
\ket{n_1,n_2} = \frac{1}{\sqrt{\kle{n_1}\,! \kle{n_2}\,!}} \klr{a_1
^{\dag}}^{n_1} \klr{a_2^{\dag}}^{n_2} \ket{0,0}\; .
\label{eq41}
\ee
to obtain the matrix representations
\be
\braket{n'_1,n'_2}{a_1 \; n_1, n_2}  =  \sqrt{\kle{n_1}} \;
\delta_{n'_1, n_1 -1} \delta_{n'_2 n_2}
\nonumber \\
\braket{n'_1,n'_2}{a_2 \; n_1, n_2} = \klr{-1}^{n_1} \;
\sqrt{\kle{n_2}} \; \delta_{n'_1 n_1} \delta_{n'_2,n_2-1}
\nonumber \\
\braket{n'_1,n'_2}{a_1^{\dag} \; n_1, n_2} = \sqrt{\kle{n_1
+1}} \;  \delta_{n'_1,n_1 +1} \delta_{n'_2 n_2}
\nonumber \\
\braket{n'_1,n'_2}{a_2^{\dag} \; n_1, n_2} = \klr{-1}^{n_1}
\sqrt{\kle{n_2 + 1}} \; \delta_{n'_1 n_1} \delta_{n'_2, n_2 +1}
\; .
\label{eq42}
\ee
In order to have finite dimensional matrices, $q$ must be a root
of unity $q=\exp\klg{\frac{2\pi i}{d}}$, implying $\kle{d}=0$
and the weakened Pauli principle $(a_\al^{\dag})^d=a_\al^d =0$.

Contrary to the Fock space theory of the non-deformed
CAR--algebra there are some important peculiarities in the
deformed case. First it is possible that $\kle{n}<0$, causing
imaginary matrix elements in (\ref{eq42}) and entailing the
necessity to represent the ${\dag}$--operation by transposition
and not Hermitean conjugation. In addition, if $d$ is even we
have $\kle{d/2}=0$ implying $a^{d/2}=0$, and furthermore next to
$\ket{0,0}$ there are additional `Fock vacua' $\ket{d/2,0},\,
\ket{0,d/2}$, and $\ket{d/2,d/2}$ all being annihilated by $a_1$
and $a_2$. As a
consequence the representation (\ref{eq42}) turns out to be
reducible.

We can simulate the vacuum representation (\ref{eq21}) by
setting $d=4$, i.e. $q=i$. We then have the Pauli principle
$a_\al^2= (a_\al^{\dag})^2=0$ and
\be
\kle{0}=\kle{2}=\kle{4}=0 \,,\quad \kle{1}=-\kle{3}=1 \;.
\label{eq43}
\ee
Defining the basis vectors
\be
\begin{array}{rclrcl}
\ket{0,0} & \Longrightarrow & \ket{\beta_{{11}}^1}
\qquad \qquad  \ket{2,2} & \Longrightarrow & \ket{\beta_{{00}}^1} \\
\ket{1,0} & \Longrightarrow & \ket{\beta_{{11}}^2}
\qquad \qquad  \ket{3,2} & \Longrightarrow & \ket{\beta_{{00}}^2} \\
\ket{0,1} & \Longrightarrow & \ket{\beta_{{11}}^3}
\qquad \qquad  \ket{2,3} & \Longrightarrow & \ket{\beta_{{00}}^3} \\
\ket{1,1} & \Longrightarrow & \ket{\beta_{{11}}^4}
\qquad \qquad  \ket{3,3} & \Longrightarrow & \ket{\beta_{{00}}^4}
\\[2ex]
\frac{1}{\sqrt{2}} \klr{\ket{0,2} + \ket{2,0}} & \Longrightarrow
& \ket{\beta_{{\frac{1}{2}0}}^1}
\qquad \qquad  \ket{3,0} & \Longrightarrow & \ket{\beta_{{\frac{1}{2}0}}^5} \\
\ket{3,1} & \Longrightarrow & \ket{\beta_{{\frac{1}{2}0}}^2}
\qquad \qquad  \ket{0,3} & \Longrightarrow & \ket{\beta_{{\frac{1}{2}0}}^6} \\
\frac{1}{\sqrt{2}} \klr{\ket{0,2} - \ket{2,0}} & \Longrightarrow
& \ket{\beta_{{\frac{1}{2}0}}^3}
\qquad \qquad  \ket{1,2} & \Longrightarrow & \ket{\beta_{{\frac{1}{2}0}}^7} \\
- \ket{1,3} & \Longrightarrow & \ket{\beta_{{\frac{1}{2}0}}^4}
\qquad \qquad  \ket{2,1} & \Longrightarrow & \ket{\beta_{{\frac{1}{2}0}}^8}
\end{array}
\label{eq44}
\ee
and
\be
A_{{11}}^{ij} = \braket{\beta_{{11}}^i}{A \,
\beta_{11}^j}\,,\quad A_{{00}}^{ij}=\braket{\beta_{{00}}^i}{A \,
\beta_{{00}}^j}\,,\quad A_{{\frac{1}{2}0}}^{ij} =
\braket{\beta_{{\frac{1}{2}0}}^i}{A \, \beta_{\frac{1}{2}0}^j}
\label{eq45}
\ee
shows that inspite of (\ref{eq43}) the sectors labeled by $11,\,
00$, and $\frac{1}{2}0$ are invariant under the application of
the deformed field operators and deformed observables. Therefore
an arbitrary algebra element decomposes into the direct sum
\be
A=A_{11}\oplus A_{00} \oplus A_{\frac{1}{2}0} \;.
\label{eq46}
\ee
If these matrix representations are compared with the matrix
representations of (\ref{eq21}) one gets the identifications
\be
\pi_{vw}(c_\al) = \frac{1}{2} \klr{a_\al+\bar{a}_\al} +
\frac{1}{2i} \klr{a_\al - \bar{a}_\al}^{\dag} \nn\\
N=\kle{N_1}+\kle{N_2}\,,\quad S_3 = \frac{1}{2}
\klr{\kle{N_2}-\kle{N_1}}
\label{eq47}
\ee
and the extremal vacua $\ket{\Om_{11}},\,\ket{\Om_{00}},\,
\ket{\Om_{\frac{1}{2}0}}$ correspond to the Fock vacua
$\ket{0,0},\, \ket{2,2}$ and $\frac{1}{\sqrt{2}}
\klr{\ket{0,2}+\ket{2,0}}$, respectively.

There is a slight deviation of the Fock representation of the
$q$--deformed CAR--algebra and the vacuum representations of the
field and observable algebra, too. Although the observables $N$
and $S_3$ can be expanded in terms of elements of the deformed
CAR--algebra, this is in general not possible for the spin
operators $S_1$ and $S_2$. We can only give suitable expressions
for $S_1$ and $S_2$ in the sectors labeled by $11$ and $00$, but
not in the $\frac{1}{2}0$ sector. Therefore we can not
distinguish the spin--1 eigenvector
$\ket{\beta^3_{\frac{1}{2}}}$ of $\vec{S}^{\,2}$ from a vacuum
vector here. In order to have the full identification of the
deformed CAR--algebra in Fock representation with the
non-deformed algebra ${\cal A}_G(\HH)$ in a vacuum
representation we must adjoin the observables $S_1$ and $S_2$ in
the deformed case.

\subsection{Thermo field dynamics}

The basic idea of thermo field dynamics is to give the
thermodynamic average at finite temperature $\left<A\right>_\beta =
tr \klr{A\, \exp\{-\zeta -\beta H\}}$ in terms of an expectation value
$\left<A\right>_\beta = \braket{\Om_\beta}{A\,\Om_\beta}$ with a
`thermal vacuum' $\ket{\Om_\beta}$ \cite{tfd}.
\nocite{takahashiumezawa75} This allows Green's function
techniques to be applied to thermal quantum fields and can be
shown to be equivalent to the algebraic setup of Haag, Hugenholz
and Winnink in equilibrium thermodynamics \cite{tfdappl}.
\nocite{matsumoto77,msut80,ojima81}

To achieve this aim the
original theory has to be embedded into an enlarged theory, which
is done by introducing new and independent tilde fields.
Restricting ourselves to the case $\HH=\C$ the basic equations
are the anticommutation relations
\be
\klg{\tilde{a},\,a} =\klg{\tilde{a}^\ast,\,a} =
\klg{\tilde{a},\,\tilde{a}} = \klg{a,\,a}=0\,,\quad
\klg{\tilde{a}^\ast,\,\tilde{a}} = \klg{a^\ast,\,a}=\1\,,\quad
\label{eq31}
\ee
given in their Fock representation with
$a\ket{0}=\tilde{a}\ket{0} =0$. According to thermo field
dynamics the thermal vacuum must be of the form
\be
\ket{\Om_\beta}
=\klr{\lam_1(\beta)+\lam_2(\beta)a^\ast\tilde{a}^\ast }\ket{0}
\label{eq32}
\ee
here. For the Hamiltonian $H=\ep_1 a^\ast a+\ep_0a\,a^\ast$ the
parameters are determined to be
\be
\lam_1(\beta)=\frac{1}{\sqrt{1+e^{-\beta
(\ep_1-\ep_0)}}}\,,\quad
\lam_2(\beta)=\frac{e^{-\frac{1}{2}\beta (\ep_1-\ep_0)}}{\sqrt{1+e^{-\beta
(\ep_1-\ep_0)}}}\;.
\label{eq33}
\ee
Further the prescription to implement the observables in thermo
field dynamics is to put
\be
\hat{N} = a^\ast a - \tilde{a}^\ast \tilde{a} \;.
\label{eq34}
\ee

Identifying the operators $c=a$, $c^\ast=a^\ast$, and
$N=\hat{N}$ shows that the subalgebra of (\ref{eq31}) generated
by $a, \,a^\ast$, and $\hat{N}$ is completely equivalent with
(\ref{eq17}) and (\ref{eq16}). Thus, the inequivalent vacuum
representations are contained as the canonical substructure
given by the original fields and observables in the Fock
representation of the enlarged algebra of thermo
field dynamics.

Furthermore, also identifying the states in terms of
\be
\om_v\klr{F(c,c^\ast)} = \bra{\Om_\beta} F(a,a^\ast)
\ket{\Om_\beta}
\label{eq35}
\ee
with an arbitrary function $F$
permits us to calculate the corresponding temperature dependency
of the parameter $v$ to be
\be
v(\beta)=\frac{1}{1+e^{-\beta(\ep_1-\ep_0)}}\;.
\label{eq36}
\ee

We will give a thermodynamic interpretation of the
parameter $v$ in a different and more systematic way now.

\section{Vacuum dynamics and the thermodynamic equilibrium state}

In this section we discuss the problem of non-trivial dynamics
connecting inequivalent representations, i.e. the dynamics of
vacuum states. We will derive a generalized Schr\"odinger
equation and give its solution for the dynamics on the vacua of
${\cal A}_{U(1)}(\C)$. Since this dynamics is irreversible we
seek to give a thermodynamic interpretation of the equilibrium
state.

Consider the space of vacuum states ${\cal V}$ with an arbitrary (finite)
number of extremal vacua $\om_r$ and extremal decomposition
$\om=\sum_r\lam_r\,\om_r$. For a given dynamics $\beta_t:{\cal V}
\rightarrow {\cal V}$ the adjoint dynamics on the algebra is
defined by
\be
\klr{\beta_t\,\om} \klr{A} = \om\klr{\al_t (A)} \;.
\label{eq48}
\ee
If $\al_t$ is unitarily implemented it follows that $\beta_t$
conserves convex combinations and since for any continous
dynamics we have $\beta_t(\om_r)=\om_r$
it is $\beta_t(\om)=\om$ for all vacuum states. Hence $\al_t$
cannot be unitarily implemented if there should be a
non-trivial vacuum dynamics at all.

\begin{theorem}
Any infinitesimally generated vacuum dynamics must
satisfy the generalized Schr\"odinger equation for the extremal
states
\be
\frac{d}{dt} \om_r\klr{\al_t(A)} = \sum_s \ga^{sr}\,
\om_s\klr{\al_t(A)} \;,
\label{eq49}
\ee
with $\ga^{ss}\leq 0$, $\ga^{sr}\geq 0$ for $s\not=r$, and
$\sum_s \ga^{sr}=0$.
\end{theorem}
\proof
We must require that $\beta_t(\om)$ is again a
vacuum, i.e. $\beta_t$ preserves linearity, positivity,
normalization, and the vacuum quantum numbers of observables.
For the state $\om=\sum_r\lam_r\,\om_r$ we can expand
$\beta_t(\om) =\sum_r\beta^r_t(\lam)\,\om_r$ in terms of the
extremal vacua again. The functions $\beta^r_t(\lam)$ must
satisfy the subsidary conditions $\beta^r_t(\lam)\geq 0$ and $\sum_r
\beta^r_t(\lam)=1$. Making use of (\ref{eq48}) with the
projector $P_s$ onto the extremal vacuum $\om_s$ yields
$\beta^s_t(\lam)=\sum_r\lam_r\,\om_r\klr{\al_t(P_s)} $
and thus the functions $\beta^s_t(\lam)=\sum_r \lam_r\,
\beta^{sr}_t$ must be linear in $\lam$, and the coefficients satisfy
$\beta^{sr}_t\geq 0$, $\sum_s \beta^{sr}_t=1$, and the initial
condition $\lim_{t\rightarrow 0}\beta^{sr}_t = \delta_{sr}$.
Inserting this into equation (\ref{eq48}) gives
\be
\sum_s \beta^{sr}_t\, \om_s(A)=\om_r\klr{\al_t(A)} \;.
\label{eq50}
\ee
If the dynamics is infinitesimally generated, i.e.
$\frac{d}{dt}\beta^{sr}_t= \sum_{r'}\ga^{sr'}\beta^{r'r}_t$, the
Schr\"odinger equation (\ref{eq49}) follows by differentiating
(\ref{eq50}). The subsidary conditions for $\ga^{sr}$ are direct
consequences of the subsidary conditions for $\beta^{sr}_t$,
since $\ga^{sr}=\lim_{t\rightarrow 0} \frac{d}{dt} \beta^{sr}_t$.
\qed
The main drawback of the Schr\"odinger equation (\ref{eq49}) is
that the dynamical matrix $\ga$ is only restricted by general
properties serving to preserve the characteristics of vacuum
states in time. For the evaluation of concrete models we should
determine $\ga$ as a function of the model Hamiltonian. However,
this is a non-trivial task going beyond conventional quantum
theory where $\ga=0$. We will approach this problem in giving
a thermodynamic interpretation of the solution of (\ref{eq49})
for the case of two extremal vacua. First notice:
\begin{lemma}
The dynamical matrix in (\ref{eq49}) can be decomposed into a
sum of dynamical matrices, each mediating only between two
extremal states and possessing the same characteristic
structure.
\end{lemma}
\proof
We make use of the subsidary condition $\ga^{ss}=
-\sum_{r\not=s}\ga^{sr}$ to eliminate $\ga^{ss}$. Then it
follows by a straightforeward calculation that
\be
\ga^{rs}= \sum_{i=2}^{n}\sum_{j=1}^{i-1} \klr{D^{ij}}^{rs}
\label{eq51}
\ee
with $\klr{D^{ij}}^{rs} = -\ga^{ij}\delta_{ir}\delta_{is} +
\ga^{ij}\delta_{ir} \delta_{js} + \ga^{ji}
\delta_{jr}\delta_{is} - \ga^{ji} \delta_{jr}\delta_{js}$
being the characteristic dynamic matrices intertwining only
the extremal states $\om_i$ and $\om_j$.
\qed
Restricting ourselves to the case of two extremal states and
putting $\ga_1=\ga^{12}$ and $\ga_2=\ga^{21}$, both being
positive, we have to solve the Schr\"odinger equation
\be
\frac{d}{dt}\klr{\begin{array}{c}\om_{1}(\al_t(A)) \\
\om_0(\al_t(A)) \end{array}} = \klr{\begin{array}{cc}
-\ga_1&\ga_1\\ \ga_2&-\ga_2\end{array}}
\klr{\begin{array}{c}\om_{1}(\al_t(A)) \\
\om_0(\al_t(A)) \end{array}}\;.
\label{eq53}
\ee
\begin{lemma}
The eigenvalues of the dynamical matrix in (\ref{eq53}) are
calculated to be $d_1=0$ and $d_2=-(\ga_1+\ga_2)\leq 0$.
\end{lemma}
Hence we expect the dynamics to asymptotically approach a
limiting state.
\begin{lemma}
The solution to the vacuum dynamics (\ref{eq53}) for the state
$\om_v = v\,\om_1+(1-v)\om_0$ is given by
\be
\om_v(\al_t(A)) = \kle{{\ts \frac{\ga_2}{\ga_1+\ga_2}} \om_1(A) +
{\ts \frac{\ga_1}{\ga_1+\ga_2}} \om_0(A)} - e^{-(\ga_1+\ga_2)t}
[{\ts \frac{\ga_2}{\ga_1+\ga_2}} - v][\om_1(A)-\om_0(A)] \,.
\label{eq54}
\ee
\end{lemma}
\proof
Introducing the functionals $\Delta=\om_1-\om_0$ and
$\Sigma=\om_1+\om_0$ we get the differential equations
$\frac{d}{dt} \Delta (\al_t(A))=-(\ga_1+\ga_2)\Delta(\al_t(A))$ and
$\frac{d}{dt} \Sigma(\al_t(A)) = (\ga_2-\ga_1) \Delta(\al_t(A))$. They are
solved by $\Sigma(\al_t(A)) = \Sigma(A) + \frac{\ga_2-\ga_1}{\ga_1+\ga_2}
\klr{1-\exp\{-(\ga_1+\ga_2)t\}}\Delta(A)$ and $\Delta(\al_t(A))
= \exp\{-(\ga_1+\ga_2)t\}\Delta(A)$. Inserting these
solutions into $\om_v= \frac{1}{2}\Sigma+(v-\frac{1}{2})\Delta$
yields (\ref{eq54}).
\qed
It is obvious from (\ref{eq54}) that the vacuum dynamics is
necessarily irreversible unless $\ga_1=\ga_2=0$. The equilibrium
state
\be
\lim_{t\rightarrow 0} \om_v(\al_t(A)) =
\klr{\frac{\ga_2}{\ga_1+\ga_2} \om_1(A) +
\frac{\ga_1}{\ga_1+\ga_2} \om_0(A)}
\label{eq55}
\ee
is independent from an arbitrarily chosen initial state and
completely determined by the ratio $\ga_1/\ga_2$.

In order to give a thermodynamic discussion of the equilibrium
state we make the following definitions:
\begin{definition}
For a given Hamiltonian $H$ the inner energy of the equilibrium
state is
\be
U=\lim_{t\rightarrow 0} \om_v(\al_t(H))
\label{eq56}
\ee
and with $v_\infty=\frac{\ga_2}{\ga_1+\ga_2}$ the entropy is
defined by
\be
S=-k\klr{v_\infty\ln v_\infty +(1-v_\infty)\ln (1-v_\infty)}\;,
\label{eq57}
\ee
i.e. the mixing parameters in the extremal decomposition are
interpreted as the eigenvalues of the equilibrium density
operator.
\end{definition}
\begin{lemma}
For the Hamiltonian $H=\ep_1\,c^\ast c +\ep_0\,c\,c^\ast$,
($\ep_1>\ep_0$) the mixing parameter $v_\infty$ and the inverse
temperature defined by
$\beta(U)=k^{-1}\frac{\partial S}{\partial U}$ result to be
\be
v_\infty (U)= \frac{\ep_1-U}{\ep_1-\ep_0}\,,\quad
\beta(U) = \frac{1}{\ep_1-\ep_0}\ln
\klr{\frac{\ep_1-U}{U-\ep_0}} \;.
\label{eq58}
\ee
\end{lemma}
\proof
The solution for $v_\infty(U)$ is a straightforeward evaluation
of (\ref{eq56}). Inserted into (\ref{eq57}) yields (\ref{eq58})
upon differentiation.
\qed
\begin{lemma}
The temperature dependency of $v_\infty$ is derived to be
\be
v_\infty(\beta) = \frac{1}{1+e^{-\beta(\ep_1-\ep_0)}}\;.
\label{eq59}
\ee
\end{lemma}
\proof
Equation (\ref{eq58}) can be resolved to give
\be
U(\beta)= \frac{\ep_0\,e^{-\beta\ep_0} + \ep_1\,
e^{-\beta\ep_1}}{e^{-\beta\ep_0} + e^{-\beta\ep_1}} \;.
\label{eq60}
\ee
Inserting this into the expression for $v_\infty(U)$ yields
(\ref{eq59}).
\qed
Notice that equation (\ref{eq59}) coincides exactly with
(\ref{eq36}). Further
we would have obtained the same inner energy and entropy
function by evaluating $U=tr(\varrho H)$ and $S=-k \, tr(\varrho
\ln \varrho)$ for the density operator
$\varrho=\exp\{-\zeta-\beta H\}$. However our results were
derived from the solution of a non-trivial vacuum dynamics here!
It was shown that any vacuum dynamics is irreversible and the
asymptotic equilibrium state can be thermodynamically
interpreted inspite of the definitions (\ref{eq56}) and
(\ref{eq57}).

Looking at the time-dependent solutions (\ref{eq54}) shows that
the equilibrium is not affected by the value of $\ga_1+\ga_2$, but
only by $\ga_1/\ga_2$. Rather, the independent value
$\ga_1+\ga_2$ characterizes the relaxation time of the system.
Considering the ratio
\be
\frac{\ga_1}{\ga_2} = \frac{\om_{v_\infty}(H) -
\om_0(H)}{\om_1(H)-\om_{v_\infty}(H)}
\label{eq61}
\ee
one might speculate that $\ga_1=\hbar^{-1} (\om_{v_\infty}(H) -
\om_0(H))$ and $\ga_2=\hbar^{-1} (\om_1(H)-\om_{v_\infty}(H))$,
implying $\ga_1+\ga_2=\hbar^{-1} (\om_1(H)-\om_0(H))$. Then the vacuum
dynamics is generated by the fact that the Hamiltonian has
different expectation values for different inequivalent vacuum
representations.

\section{Application to BCS--theory}

The conventional solutions of BCS--theory either make use of a
suitably chosen ansatz for the ground state wave function or
reformulate the theory in terms of an `equivalent' Hamiltonian
which is only quadratic in the field operators and permits exact
solutions in order $O(1/N)$ \cite{bcs}.
\nocite{bcs57,mattislieb,bogoliubov}
Anyhow the resulting ground state
of the theory is not an eigenstate of the particle number
operator. In the first case this originates from the ansatz
itself and in the second case from the fact that the
`equivalent' Hamiltonian does not commute with the particle
number operator where as the original BCS--Hamiltonian did.

We will show here that these solutions are based on an
inconsistency when formulated in Fock space. Admitting
non-trivial vacua we will obtain exact particle number invariant
solutions to the full BCS--Hamiltonian
\be
H=\sum_z \ep_z\,c^\ast_zc_z -\frac{1}{2} \sum_{zz'}
\frac{V_{zz'}}{\hat{N}}
c^\ast_{z'}c^\ast_{\hat{z}'} c_zc_{\hat{z}} \;,
\label{eq62}
\ee
where $z=({k},\al)$, $\hat{z}=(-{k},-\al)$, and
$\hat{N}$ is the (finite) number of lattice points. The
potential is taken to have the symmetry properties
$V_{zz'}=V_{z'z}=-V_{\hat{z}z'}$.

Considering the averaged operator
\be
B^\ast_z = \sum_{z'} \frac{V_{zz'}}{\hat{N}}
c^\ast_{z'}c^\ast_{\hat{z}'}
\label{eq63}
\ee
one can show that $B^\ast_z$ commutes with every field operator
in order $O(1/\hat{N})$. If in the limit
$\hat{N}\rightarrow\infty$ the theory were still represented in an
irreducible representation --- like for instance the Fock
representation --- it follows that $B^\ast_z=\Delta_z \1$ with
$\Delta_z\in\C$.  On the other hand the commutator with the
particle number operator $\kle{B^\ast_z,\,N}=-2B^\ast_z$ is not
affected by taking the thermodynamic limit. Since $\kle{\1,N}=0$
we have $B^\ast_z=0$ in the thermodynamic limit. Hence, either
the theory is not represented in an irreducible representation
in the thermodynamic limit or it must be the theory of free
fields and there is no superconductivity.

To get a grasp of the theory with infinitely many degrees of
freedom from its finite counterpart we will consider vacuum
states on the algebra defined by the relations
\be
\klg{c_z,\,c_{z'}}=0\,,\quad \klg{c_z^\ast,\,c_{z'}}=
\delta_{zz'}\1\,,\quad \kle{N,\,c_z}=-c_z\,, \nn\\
\kle{S^k,\,c_{{k}\al}}= \sum_{\al'}\sig^k_{\al\al'} c_{k\al'}
\,,\quad U_nc_{k\al}U^\ast_n=e^{ikn}c_{k\al} \;.
\label{eq64}
\ee
The operators $U_n$ are the implemented translation
operators. Since any vacuum state satisfies
$\om(AN)=\om(AS^k)=0$ and $\om(AU_n)=\om(AU^\ast_n)=\om(A)$ we
conclude as before that $\om(c^\ast_{z_1}\ldots
c^\ast_{z_n}c_{z'_1} \ldots c_{z'_{n'}})$ is zero unless $n=n'$,
$\sum_{i}(\al_i-\al_i')=0$ and $\sum_i (k_i-k_i')\in 2\pi\Z$.

\begin{theorem}
The complete hierarchy of non-zero expectation values, i.e. the
non-trivial vacuum state of the theory, can be
determined in order $O(1/\hat{N})$ from the BCS--Hamiltonian
(\ref{eq62}) and the following assumptions:
\begin{itemize}
\item The condensate equation: $\om(B^\ast_zB_{z'}A) = \om
      (B^\ast_zB_{z'})\,\om(A)$
\item The equilibrium of the condensate: $\kle{B_z,\,H}=0$\,.
\end{itemize}
\end{theorem}
{\bf Comment:}
Notice that the operators $B^\ast_zB_{z'}$ are in the center of
the complete field and observable algebra in order
$O(1/\hat{N})$, but they must not be proportional to unity.
Rather, the condensate equation is a requirement on the vacuum
{\it state}. Since the operators $B_z$ will determine the vacuum
solutions we pick out the equilibrium vacuum by the second
assumption. We will not investigate non-equilibrium
properties of the BCS model here.

\noindent\proof
Defining the generating functional of time-ordered
$\om$--functions with completely anticommuting sources $\eta$
and $\bar{\eta}$
\be
G(\eta,\bar{\eta})&=&\sum_{n=0}^\infty (n!)^{-2}\hspace{-1.5em}
\sum_{{\scriptstyle \vspace*{-0.5ex} \begin{array}{c} z_1\ldots
z_n\\[-0.5ex]z'_1\ldots z'_n
\end{array}}} \hspace{-1em} \int dt_1\ldots dt_ndt'_1\ldots
dt'_n \, \eta_{z_1}(t_1) \ldots \eta_{z_n}(t_n)
\bar{\eta}_{z'_1}(t'_1)  \ldots \bar{\eta}_{z'_n}(t'_n) \times
\nn\\[-4ex]
&&\hspace{9em}\times \om({\cal T} c^\ast_{z_1}(t_1) \ldots
c^\ast_{z_n}(t_n) c_{z'_1}(t'_1) \ldots c_{z'_n}(t'_n))
\label{eq65}
\ee
yields the functional equations of motion
\be
i\frac{\partial}{\partial t} \abl{\eta_{z}(t)}
G(\eta,\bar{\eta}) = \kle{-\bar{\eta}_z(t) + \ep_z
\abl{\eta_{z}(t)} + D_z \abl{\bar{\eta}_{\hat{z}}(t)}}
G(\eta,\bar{\eta}) \nn\\
i\frac{\partial}{\partial t} \abl{\bar{\eta}_{z}(t)}
G(\eta,\bar{\eta}) = \kle{-\eta_z(t) - \ep_z
\abl{\bar{\eta}_{z}(t)} - \bar{D}_z \abl{\eta_{\hat{z}}(t)}}
G(\eta,\bar{\eta})
\label{eq66}
\ee
where the abbreviations
\be
D_z = \sum_{z'} \frac{V_{zz'}}{\hat{N}}
\abl{\eta_{z'}} \abl{\eta_{\hat{z}'}} \;,\quad  \bar{D}_z = \sum_{z'}
\frac{V_{zz'}}{\hat{N}}
\abl{\bar{\eta}_{z'}} \abl{\bar{\eta}_{\hat{z}'}}
\label{eq67}
\ee
have been used. Due to the equilibrium of the condensate the
functional operators $D_z$ and $\bar{D}_z$ are time independent
in (\ref{eq66}).  Differentiating (\ref{eq66}) once more with
respect to $t$, defining $\Delta^2_z = \om (B^\ast_zB_{z})$, and
making use of the condensate equation and of (\ref{eq66}) we obtain
\be
- \kle{\frac{\partial^2}{\partial t^2} +\ep_z^2+\Delta_z^2}
\abl{\eta_{z}(t)} G(\eta,\bar{\eta}) &=&
\kle{-\klr{i\frac{\partial}{\partial t} + \ep_z} \bar{\eta}_z (t) -
D_z\eta_{\hat{z}} (t)}  G(\eta,\bar{\eta}) \nn \\
- \kle{\frac{\partial^2}{\partial t^2} +\ep_z^2+\Delta_z^2}
\abl{\bar{\eta}_{z}(t)} G(\eta,\bar{\eta}) &=&
\kle{-\klr{i\frac{\partial}{\partial t} - \ep_z} \eta_z (t) -
\bar{D}_z \bar{\eta}_{\hat{z}} (t)}  G(\eta,\bar{\eta}) \;.
\label{eq68}
\ee
Since we are interested in the hierarchy of equal-time
$\om$--functions the equal-time limit of (\ref{eq68}) is needed.
Inspite of the time derivatives present in (\ref{eq68}) the
equal-time limit will be taken in the following way: Comparing
the coefficients of the functionals in (\ref{eq68}) we get
differential equations for the
coupled hierarchy of multi-time $\om$--functions. These
equations are then Fourier transformed and subsequently all
Fourier variables
are integrated over. Whenever necessary the poles on the real
axis are circumvented by replacing
$\ep^2_z+\Delta^2_z \Rightarrow \ep^2_z+\Delta^2_z\pm i\delta$ and
taking the limit $\delta\rightarrow 0$ at the end. Collecting
the resulting
equal-time hierarchy equations by a functional equation again
yields
\be
\abl{\eta_z} A(\eta,\bar{\eta}) &=&
\kle{\frac{-\ep_z}{2E_z}\bar{\eta}_z + \frac{1}{2E_z} D_z
\eta_{\hat{z}}}  A(\eta,\bar{\eta}) \nn\\
\abl{\bar{\eta}_z} A(\eta,\bar{\eta}) &=&
\kle{\frac{+\ep_z}{2E_z}\eta_z + \frac{1}{2E_z} \bar{D}_z
\bar{\eta}_{\hat{z}}}  A(\eta,\bar{\eta})\;,
\label{eq69}
\ee
where $E_z = \sqrt{\ep_z^2+\Delta_z^2}$, $A(\eta,\bar{\eta})$
denotes the generating functional of antisymmetric equal-time
$\om$--functions, and we have put $\eta_z=\eta_z(t=0)$ and
$\bar{\eta}_z = \bar{\eta}_z(t=0)$.

The solution of (\ref{eq69}) is simplified by the ansatz
\be
A(\eta,\bar{\eta}) = \exp \klr{-\sum_z \eta_z \frac{\ep_z}{2E_z}
\bar{\eta}_z}  A'(\eta,\bar{\eta})
\label{eq70}
\ee
implying
\be
\abl{\eta_z} A'(\eta,\bar{\eta}) = \frac{1}{2E_z} D_z
\eta_{\hat{z}} A'(\eta,\bar{\eta}) \,,\quad
\abl{\bar{\eta}_z} A'(\eta,\bar{\eta}) = \frac{1}{2E_z} \bar{D}_z
\bar{\eta}_{\hat{z}} A'(\eta,\bar{\eta}) \;.
\label{eq71}
\ee
Setting $\Delta_{zz'}=\om (B^\ast_zB_{z'})$
it follows that $\Delta^2_{zz'}=\Delta^2_z\Delta^2_{z'}$ in order
$O(1/\hat{N})$. Furthermore defining
\be
{\cal N} = \sum_z \eta_z \abl{\eta_z} \,,\quad \bar{{\cal N}} =
\sum_z \bar{\eta}_z \abl{\bar{\eta}_z} \,,\quad  {\cal K} =
\sum_{zz'} \frac{\Delta_z\Delta_{z'}}{4E_zE_{z'}}
\bar{\eta}_{z'} \eta_z \eta_{\hat{z}}\bar{\eta}_{\hat{z}'}
\label{eq72}
\ee
we get ${\cal N}\bar{{\cal N}} A' = {\cal K} A'$ from
(\ref{eq71}) and the condensate equation, which is solved
by
\be
A'(\eta,\bar{\eta}) = \sum_{n=0}^\infty \kle{(2n)!!}^{-2}
\kle{{\cal K}(\eta,\bar{\eta})}^n \;.
\label{eq73}
\ee
Since any algebra element of the CAR--algebra can be expanded in
terms of antisymmetrized field operator products the vacuum
state is completely determined by (\ref{eq70}) and (\ref{eq73}).
\qed
\begin{lemma}
The 2--point and 4--point functions of the solution found in
theorem \thetheorem\ are calculated to be
\be
\om(c^\ast_{z_1}c_{z'_1}) &=&
\frac{1}{2}\kle{1-\frac{\ep_{z_1}}{E_{z_1}}} \delta_{z_1z_1'} \nn\\
\om(c^\ast_{z_1}c^\ast_{z_2}c_{z_1'}c_{z_2'}) &=& \frac{1}{4}
\kle{1-\frac{\ep_{z_1}}{E_{z_1}}}
\kle{1-\frac{\ep_{z_2}}{E_{z_2}}} \kle{\delta_{z_2z_1'}
\delta_{z_1z_2'} - \delta_{z_1z_1'} \delta_{z_2z_2'}} +
\nn\\
&& +\frac{1}{4} \frac{\Delta_{z_1} \Delta_{z_1'}}{4E_{z_1}E_{z_1'}}
\delta_{z_1\hat{z}_2} \delta_{z_1'\hat{z}_2'} \,.
\label{eq74}
\ee
\end{lemma}
\begin{lemma}
The consistency of the condensate equation with the
4--point function of lemma \addtocounter{lemma}{-1}\thelemma\
yields the gap equation \addtocounter{lemma}{1}
\be
\Delta_z = (\pm) \frac{1}{2} \sum_{z'} \frac{V_{zz'}}{\hat{N}}
\frac{\Delta_{z'}}{2E_{z'}}
\label{eq75}
\ee
in order $O(1/\hat{N})$.
\end{lemma}
\proof
Inserting (\ref{eq74}) into
$\Delta_z\Delta_{z'}=\om(B^\ast_zB_{z'})$ gives
\be
\Delta_z\Delta_{z'}= \sum_y
\frac{V_{zy}V_{z'y}}{4\hat{N}^2}\kle{1-\frac{\ep_y}{E_y}}^2 +
\frac{1}{4} \sum_{yy'} \frac{V_{zy}V_{z'y'}}{\hat{N}^2}
\frac{\Delta_y\Delta_{y'}}{4E_yE_{y'}}\,.
\label{eq76}
\ee
The first term on the right hand side vanishes in order
$O(1/\hat{N})$ implying the squared gap equation (\ref{eq75}).
\qed
One can continue to build up the complete BCS--theory, the only
remarkable difference being the fact that the state $\om$ is a
true vacuum state with respect to the full field operator and
observable algebra here. It is the exact, particle
number invariant  ground state of
the BCS--Hamiltonian (\ref{eq61}) in order $O(1/\hat{N})$.

\section{Conclusions}

If the algebra of fields is enlarged in a canonical way by
adjoining the observables as independent operators there exist
inequivalent (vacuum) representations for quantum theories with
finitely many degrees of freedom, too.

These inequivalent
representations provide the basis for the consistent description
of a large variety of collective microscopic and macroscopic
phenomena in many-particle quantum theory. Some of their
properties can be recovered from quasi-particle methods in Fock
space, but for one thing these need not exhaust the complete
structure and in addition might yield inconsistencies in the
thermodynamic limit. Nevertheless, these methods have proved to
be  successful in many physical applications. Therefore,
inequivalent representations are expected to play a decisive
role in finite nuclear shell models, in finite lattice models,
etc., where these considerations have so far been considered
only indirectly. If applied to infinite systems one can approach
its solution consistently from its finite solutions, as was
demonstrated in the case of the BCS--model.

The possibility of non-trivial vacua is closely engaged with the
thermodynamics of many-particle quantum field theory. If the extremal
decomposition of the equilibrium vacuum state is known, all
thermodynamic functions can be calculated straightforeward and
without an additional maximum principle of the entropy. The
irreversible vacuum dynamics resulting from the Schr\"odinger
equation for the extremal vacuum states might serve to
microscopically investigate non-equilibrium properties in
many-particle physics, if it is possible to determine the
dynamical matrix $\ga$ by the model Hamiltonian.

\end{document}